\documentclass[twocolumn,showkeys,preprintnumbers,amsmath,amssymb]{revtex4}
\usepackage{graphicx}

\usepackage{fancyhdr}
\begin{document}

\title{STUDY OF THE COMPOSITION OF PRODUCTS OF CONTROLLED NUCLEOSYNTHESIS
BY LOCAL AUGER-ELECTRON SPECTROSCOPY}

\author{S. V. Adamenko}
\author{A. S. Adamenko}
\author{S. S. Ponomarev}
\altaffiliation{Institute of Problems of Materials Science of the
NAS of Ukraine, 3 Krzhizhanovsky Str., Kyiv, 03142 Ukraine}

\affiliation{%
Electrodynamics Laboratory ``Proton--21''\\
14/1 Dovjenko Str., 03057, Kiyv, Ukraine}%
\email{enr30@enran.com.ua}

\date{\today}

\begin{abstract}%
By local Auger-electron spectroscopy on solid targets and
accumulating screens, we studied the composition of
nucleosynthesis products, in which we expected to reveal the
presence of long-lived transuranium elements (LTE). The goal of
the work was to find LTE on the surface of the specimens under
investigation. The specificity of the surface composition
determination consisted in that the registered spectra included
usually the Auger-peaks of a lot of chemical elements, which poses
major difficulties in their identification. To solve the latter
problem, we used a wide range of energies (from 30 to 3000\,eV)
with the purpose to cover a maximum number of series of the
Auger-peaks of analyzed elements and long exposures (up to 3\,h)
to reveal Auger-peaks with small intensity. In a number of cases
for analyzed elements in complicated spectra the Auger-spectra of
corresponding pure elements or their simple compounds were
registered with a high signal-to-noise ratio. As artifacts of the
analysis, we consider such phenomena as the electric charging,
characteristic losses of energy, and chemical shift. On the study
of the surface of specimens, we found the unidentifiable
Auger-peaks with energies of 172, 527, 1096, 94, and 560\,eV and
the doublet of peaks with energies of 130 and 115\,eV. We failed
to refer them to any Auger-peaks of chemical elements in the
atlases and catalogs or to any artifacts. As one of the variants
of interpretation of the revealed peaks, we consider the
assumption about their affiliation to LTE.
\end{abstract}


\maketitle \thispagestyle{fancy}


\section*{INTRODUCTION}
 In the present work, we study the specimens whose surfaces contain
the products of the explosion-induced dispersion of targets
undergone the high-energy impact compression which was realized
with the purpose to reach the extremal density of a substance and,
presumably, to initiate the collective processes of nuclear
transformations (nucleosynthesis) terminating in the formation of
highly stable nuclei (stable and long-lived isotopes) of various
chemical elements with broad mass spectrum including transuranium
ones~\cite{1,2}.

As specimens, we take the solid targets, in which the reaction of
nucleosynthesis ran, and accumulating screens, on which the target
material undergone a nuclear transmutation is deposited after its
explosion. The products remained in the target crater and
deposited on accumulating screens are drops, splashes, films,
particles, and other micro- and nanoobjects with complicated
morphology which are irregularly distributed on their surfaces. In
the study of the reaction products, we were interested, first of
all, in the determination of their element and isotope
compositions. It is especially worth to note that we expect to
reveal the presence of LTE.

The isotope composition of the nucleosynthesis products was
carried out by laser mass-spectrometry (LMS) and secondary ion
mass-spectrometry (SIMS). The results of these investigations are
far beyond the frame of the present work and now are prepared as
separate publications. As for the problem of determination of
their element composition, we used the complex of mutually
supplementing physical methods rather than a sole one. On the one
hand, these methods must have high sensitivity and, on the other
hand, must cover the range of typical sizes of all analyzed
objects (from several mm to several nm). In fact, the analyzed
chemical elements, in particular LTE, can be distributed uniformly
in the object under study with low concentration in it or, on the
contrary, be concentrated in small particles and/or thin submicron
films on the specimen surface and on the inner interfaces between
elements of the structure which compose the surface layer. It is
also clear that the employed methods must overlap the range of
thicknesses of the studied layers since the analyzed elements can
be distributed irregularly throughout depth.

We chose local Auger-electron spectroscopy (AES) as the method for
quantitative determination of the composition of microobjects (the
objects with size more than 1\,$\mu$m were studied by X-ray
electron probe microanalysis) which were contained in the
nucleosynthesis products and had at least one submicron size (the
surface proper, inner interfaces, submicron particles and films).
Such a choice was stipulated by that the method well satisfies the
requirements given above. First of all, AES is a nondestructive
method for quantitative determination of the composition, and the
specimens being studied with it can be investigated with other
methods. The method possesses a high spatial locality
(100$\dots$50\,nm), a small depth of the analyzed region
(1$\dots$2\,nm), a wide range of registered elements (all except
for H and He), and a quite high sensitivity (0.1$\dots$1\,at.\,\%
). If one uses the method of ion etching, AES also allows one to
study the distribution of elements throughout
depth~\cite{3,4,5,6,7,8,9}.

Thus, the goal of the present work is to study the element
composition of the nucleosynthesis products by AES and to
establish the presence or the absence of LTE.

\section{MATERIALS AND METHODS}

 We determined the composition of the specimen surface with an
Auger-microprobe JAMP--10S (JEOL, Japan). Spectra were registered
at the accelerating voltage of the electron probe which was equal
to 10\,kV, a beam current of $10^{-6}\dots10^{-8}$\,A, and the
residual pressure of $5\times10^{-7}$\,Pa in a specimen chamber.
By considering such an artifact as characteristic energy losses,
we used the accelerating voltages of 5 and 3\,kV. The energy range
of a semicylindical mirror energy-analyzer of the
Auger-spectrometer was 30$\dots$3000\,eV and its energy resolution
was from 0.5\,\% to 1.2\,\%. All spectra were registered in
differential form. In the quantitative analysis, we used the
standard program for the computation of the concentrations of
elements given by the JEOL firm-producer.

As the objects of our investigation, we took the targets which
were made of light, medium, and heavy chemically pure metals with
atomic masses from 9 to 209, as well as the accumulating screens
made of chemically pure Cu, Ag, Ta, Au, and Pb. The screens were
disks from 0.5 to 1\,mm in thickness and from 10 to 15\,mm in
diameter and served as a substrate. On one of the screen surfaces,
the studied layer of a material consisting of the nucleosynthesis
products was located. The layer had a slightly pronounced relief
and the axial symmetry. Its area varied from 1 to 2\,cm$^{2}$. In
the determination of the surface composition, we used as-received
specimens, i.e., the analyzed surface was not undergone to any
cleaning procedure.

\section{RESULTS AND DISCUSSION}

 Up to now, we have studied more than 100 specimens by AES. In
all the cases (including those where the target and an
accumulating screen were made of one chemical element maximally
purified from impurities), we registered from twenty to thirty
chemical elements in specimens. The amounts of these elements
exceeded the total content of impurities in the initial materials
by several orders of magnitude. As usual, the specimens under
study contained mainly chemical elements with small and medium
atomic numbers. The fraction of heavy chemical elements is lower,
but it increases with the atomic number of a target. Experimental
data derived by several methods (including AES) are presented
in~\cite{1} by the example of the comparative analysis of the
registered composition of nucleosynthesis products on a target and
an accumulating screen made of pure copper. At present, we have a
huge database (about 800 specimens and 14500 analyses) on the
composition of the products of laboratory nucleosynthesis which
was obtained by the methods of AES, SIMS, X-ray electron probe
microanalysis, etc. It is obvious that there is no point in giving
one or several lists of the results of analysis of the
nucleosynthesis products derived by AES. Moreover, the analysis of
this database and the construction of correlation dependencies
between the compositions of products, targets, accumulating
screens and the parameters of the technological process go beyond
the frame of this work. Below, we will not concentrate on the
presence of well-known chemical elements in the analyzed specimens
and will pay attention to the attempt to answer the question about
the content of LTE in the nucleosynthesis products.

The specificity of determination of the composition of a material
layer surface under study was defined by the fact that the
registered Auger-spectra contained, as usual, the Auger-peaks of
at least 10 chemical elements. Moreover, the collection of these
elements can considerably vary not only from one specimen to
another but from one analysis point to another in the limits of
the same specimen, and the content of elements changed in a wide
range. This yields that the spectra contain a great number of
peaks which frequently overlap one another. This circumstance
hampers the identification of the registered Auger-peaks.

There was a number of other factors concerning the problem of
identification of the registered Auger-peaks. As usual, we
registered Auger-spectra in the energy range 30\dots3000\,eV. The
choice of such a wide range was caused by two reasons. On the one
hand, this choice allows us to cover the maximum number of series
of the Auger-peaks of analyzed elements and must promote solving
the problem of their identification in the low-energy region. On
the other hand, it is necessary to answer the question about the
presence of LTE in specimens, because LTE being heavy have a lot
of peaks in the high-energy region. However, the known atlases and
catalogs of Auger-electron spectra (e.g.,~\cite{10,11,12}) usually
used in the identification of registered Auger-peaks contain the
extremely limited volume of data for the energy range
1000\dots2000\,eV and no data for the range 2000\dots3000\,eV.

Discussing the problem of identification of Auger-peaks, we should
like to indicate one more circumstance. The sensitivity of the
method of Auger-electron spectroscopy in the determination of the
amount of heavy elements is less than that for light
elements~\cite{3,4,5}. Therefore, with the purpose to reveal small
amounts of the required elements and to enhance the
signal-to-noise ratio in the registered spectra, we used long
exposures (up to 3\,h) and large currents of the primary beam of
electrons (up to $10^{-6}$\,A). However, the reference catalogs of
Auger-electron spectra~\cite{10,11,12} which are used upon the
identification of Auger-peaks give a clear preference to high
energy resolution (the registration was realized on narrow slits
of spectrometers) rather than to large signal-to-noise ratio. In
other words, Auger-peaks with low intensity were sacrificed by the
reference literature to the data on a fine structure of analytical
Auger-peaks.

Therefore, as a result of the above-mentioned circumstances, a
number of Auger-peaks absent in the reference catalogs were
registered by us in the specimens under study. The identification
of each of these peaks was carried out according to the following
scheme. For the Auger-spectrum where a unidentifiable peak was
observed, we determined the collection of elements whose
Auger-peaks were present in such an Auger-spectrum. For each of
these elements, we registered its reference Auger-spectrum in the
energy range 30\dots3000\,eV with high signal-to-noise ratio (the
large duration of the exposure) with the purpose to reveal the
low-intensity Auger-peaks of the chosen element. The reference
Auger-spectra were registered on specimens made of relevant pure
elements or their simple compounds. Then we made attempt to refer
the unidentifiable peak to some low-intensity Auger-peak revealed
in the reference Auger-spectra of the chemical elements present in
the analyzed spectrum. In such a way, we identified, for example,
the low-intensity Auger-peaks of series KLL of Si and Al with
energies, respectively, of 1737 and 1485\,eV.

Some unidentifiable peaks were referred to artifacts of the
analysis. As artifacts, we considered such phenomena as the
electric charging, characteristic losses of energy, and chemical
shift. The criterion for the unidentifiable peak to be considered
as a peak of characteristic losses of energy was its shift or
disappearance upon a change in the accelerating voltage of the
primary electron beam. The criterion of appearance of the
unidentifiable peaks as a result of the electric charging was
unproper energetic position of the Auger-peaks of chemical
elements contained in the analyzed Auger-spectrum and the lack of
reproducibility of its energetic position upon a change of high
voltage and current of primary electron beam. While analyzing a
chemical shift, we took into account, first of all, its value and
the chemical environment of the analyzed peak.

After the implementation of all the above-presented stages of the
analysis, we registered 6 unidentifiable peaks. This amount was
accumulated to the time of writing the present work after the
study of the composition of surface layers of more than 100
specimens. The brief information about them is systematized in
Table~\ref{tab1}.

\renewcommand{\arraystretch}{1.2}

\begin{table*}
\centering%
\caption{Data on unidentifiable peaks in the registered Auger
electron spectra.} {\footnotesize
\begin{tabular}{|c|c|c|c|c|c|}
\hline
Peak    &Typical     &Registration  &Behavior &Specimen & Number \\[-0.9ex]
energy,  &element      &place         &in time and   &   &of obser-\\[-0.9ex]
eV       &environment  &              &under a probe &   &vations \\
\hline \hline
172         &  Si, K, Ca, Na,     &matrix     &relatively &À132, À137,     &11\\[-0.9ex]
            &  Cu, Zn, O, N,      &            &stable   &4961, 5239,     &  \\[-0.9ex]
            &  C, S, Cl           &            &             &5292            &  \\

1095-1098    &  Si, K, Ca, Na,     &matrix     &stable   &À132, À137,     &11\\[-0.9ex]
            &  Cu, Zn, O, N,      &            &             &4961, 5239,     &  \\[-0.9ex]
            &  C, S, Cl           &            &             &5292            &  \\

525-528     &  Si, Al, Ca, O, C,  &particles     &Ca grows &4169, 4540 &3 \\[-0.9ex]
           &  S, Cl, Cu, Zn       &            & in time      &                &  \\

130 (115)   &  Al, O, C, N,       &matrix     &the intensity &5633            &2 \\[-0.9ex]
            &  S, P, Cl, Cu,      &            &decreases       &                &  \\[-0.9ex]
            &  Sn, Ce             &            &             &                &  \\

94          &  C, O, Cu           &particles  &stable   &6215,            &10\\[-0.9ex]
            &                     &at about     &             &shell    &  \\[-0.9ex]
            &                     & 1\,$\mu$m depth &             &fragment   &  \\

560         &  C, O, Cu           &particles     &the intensity &6754            &1 \\[-0.9ex]
            &                     &            &decreases       &                &  \\
\hline
\end{tabular}\label{tab1}}

\end{table*}

\begin{figure}
\centering
\includegraphics[width=8 cm]{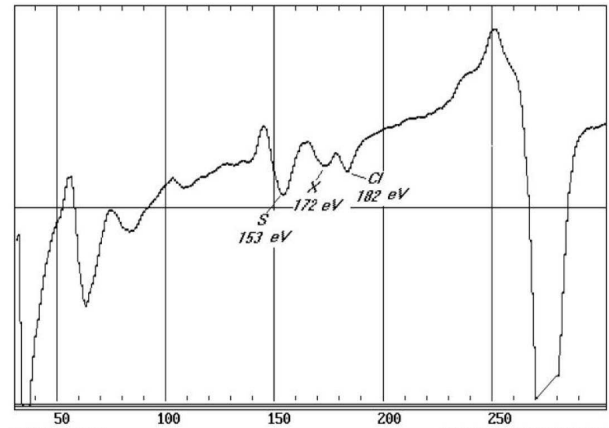}
\caption{Fragment of a typical Auger-spectrum containing the
unidentifiable peak with an energy of 172\,eV.}
\label{Auger_1.eps}
\end{figure}

 The unidentifiable peak with an energy of 172\,eV was registered
on 5 specimens. It is contained in 11 Auger-spectra registered on
these specimens. A fragment of one of these Auger-spectra
containing the unidentifiable peak with an energy of 172\,eV is
shown in Fig.~~\ref{Auger_1.eps}. The number of such spectra, by
one's wish, can be increased to any value, because the part of the
surface of each specimen where this peak can be registered is
rather extended. The peak was registered on the metal matrix
surface. Its intensity is small in all the cases and decreases
from the specimen center to its periphery. The energy position of
the peak is reproduced quite exactly. Its typical chemical
environment is Si, S, Cl, K, C, Ca, N, O, Cu, Zn, and Na, whereas
nonmetallic elements except for C and O are present usually in
slight amounts.

\begin{figure*}
\centering
\includegraphics[width=16 cm]{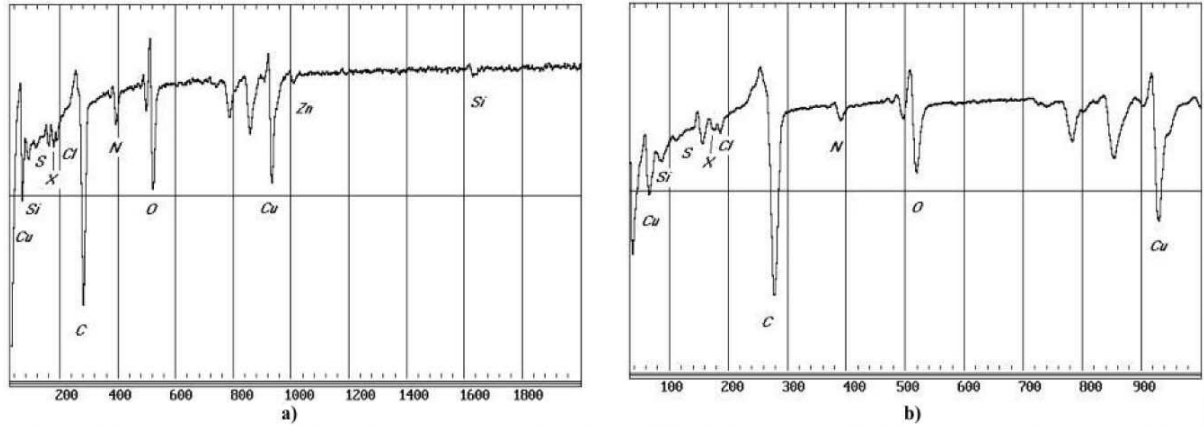}
\caption{Typical Auger-spectrum containing the unidentifiable peak
with an energy of 172\,eV (a) and the spectrum registered on the
same place in a month (b).\label{Auger_2.eps}}
\end{figure*}

We say a few words about the stability of the peak. After the
registration of the Auger-spectrum, if we at once detect it once
more on the same place, the peak intensity decreases usually by
$20\dots30$\,\%, whereas the repeated subsequent registrations
reveal no decrease in its intensity. We note that, in this case,
no significant change in the intensity of other peaks, in
particular of the Auger-peaks of C and O, occurs. That is, the
mentioned phenomenon cannot be referred to the presence of a
carbon deposit and looks as if the substrate of this peak is
decomposed in the subsurface layers under the action of the
electron beam of the probe, where this action is most intense, and
remains stable at large depths because of the decrease in the
level of action of the beam. The intensity of this peak is also
decreased in time, which is demonstrated by
Fig.~\ref{Auger_2.eps}. We present 2 Auger-spectra which contain
the unidentifiable peak with an energy of 172\,eV and were
registered on the same place in 1 month one after another: (a) and
(b), respectively. In the later case (b), the peak intensity is
clearly less than that in case (a).

 It is necessary to indicate that the unidentifiable peak with
an energy of 172\,eV appears always against the background of the
Auger-peaks of S and Cl with energies of 152 and 181\,eV,
respectively. Nevertheless, we do not refer the mentioned peak to
such an artifact of the analysis as chemical shift. Indeed, the
Auger-peaks of S and Cl are present, on the one hand, in spectra
in all the cases. On the other hand, even if we interpreted the
unidentifiable peak as the chemically shifted Auger-peak of Cl,
this shift would have an unlikely huge value (see
Table~\ref{tab2}, where we give the data on most significant known
chemical shifts of the Auger-peaks of S and Cl in their
compounds).

\begin{table}[ht]
\centering%
\caption{Data on the chemical shifts of the Auger-peaks of S and
Cl in their compounds~\cite{12}.\vspace*{1pt}} {\footnotesize
\begin{tabular}{|c|c|c|c|c|c|}
\hline
Com-&$E(\mathrm{Cl})_\mathrm{LVV}$,&$\Delta E(\mathrm{Cl})_\mathrm{LVV}$,&Com-&$E(\mathrm{S})_\mathrm{LVV}$,&$\Delta E(\mathrm{S})_\mathrm{LVV}$,\\[-0.9ex]
pound&eV            &eV                       &pound&eV           &eV \\
\hline \hline
LiCl       &179&-2&PbS      &149  &-3  \\
NaCl       &182&1 &Ag$_{2}$S&148  &-4  \\
CuCl$_{2}$ &178&-3&US$_{x}$ &147.5&-4.5\\
KCl        &178&-3&         &     &    \\

\hline
\end{tabular}\label{tab2}}
\end{table}

 Chronologically, the unidentifiable peak with an energy of 172\,eV
 was the first to be observed. Because we expected the presence
of LTE in the specimens and the main goal consisted in their
finding, we assumed that the detected peak could be one of the
brightest characteristic Auger-peaks of some LTE which is present
at the analyzed point of the specimen surface in a small amount.
Then the Auger-spectrum could contain other Auger-peaks of some
series of the same LTE with lower intensity. In other words, it
was necessary to thoroughly test the entire energy range of the
Auger-spectrum under consideration for the presence of
low-intensity peaks.

\begin{figure}[h]
\centering
\includegraphics[width=8 cm]{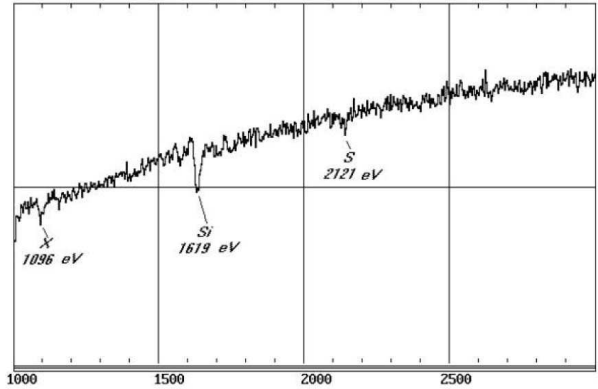}
\caption{High-energy part of the Auger electron spectrum, in the
low-energy part of which the unidentifiable peak with an energy of
172\,eV is present; the exposure duration was 3\,h. The spectrum
contains the unidentifiable peak with an energy of 1096\,eV which
regularly accompanies that with an energy of
172\,eV.\label{Auger_3.eps}}
\end{figure}

 The ``first-shot'' registration of Auger-spectra with large exposure
duration on those places of the specimen surface, where the
unidentifiable peak with an energy of 172\,eV was found, showed a
new unidentifiable peak with an energy of about 1096\,eV. A
high-energy fragment of one of these Auger-spectra is presented in
Fig.~\ref{Auger_3.eps}. We emphasize that the new peak was
registered in all cases, without exception, where the 172-eV peak
was observed (see Table~\ref{tab1}), i.e., they are observed as a
couple. By virtue of the weak intensity of the 1096-eV peak, its
exact energy position cannot be point out due to the background
fluctuations. Approximately, we may say about the interval from
1095 to 1098\,eV. The same reasons hamper the study of its
behavior in time or under the action of the electron beam of the
probe, but it is clear that the new peak is relatively stable.

\begin{figure}[h]
\centering
\includegraphics[width=8 cm]{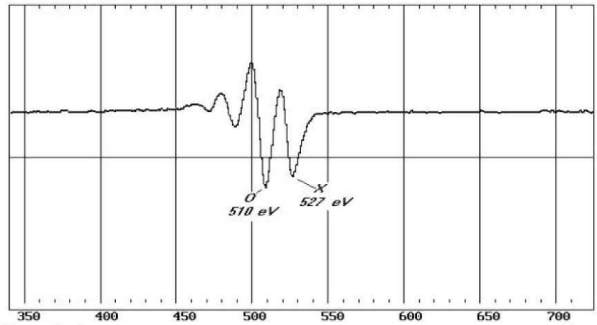}
\caption{Fragment of the Auger-spectrum containing the
unidentifiable peak with an energy of 527\,eV.\label{Auger_4.eps}}
\end{figure}

The next unidentifiable peak has energy in the the interval from
525 to 528\,eV. A fragment of the Auger-spectrum containing this
peak is given in Fig. 4. This peak was registered in 3
Auger-spectra on 2 specimens (see Table~\ref{tab1}). In all the
cases, it was found on the surface of globular particles of the
second phase (inclusions) of $50\dots70\,\mu$m in diameter. Its
intensity varied from a low to considerable one (see
Fig.~\ref{Auger_4.eps}). The peak appeared in the typical chemical
environment: Si, S, Cl, C, Ca, O, Cu, Zn, and Al. The data on the
chemical shift of the Auger-peak of O are presented in
Table~\ref{tab3}.

\begin{table}[h]
\centering%
\caption{Data on the chemical shift of the Auger-peak of O in its
compounds~\cite{12}.\vspace*{0.5pt}} {\footnotesize
\begin{tabular}{|c|c|c|c|c|c|}
\hline
&E(O)&$\triangle$E(O)&&E(O)&$\triangle$E(O)\\[-0.9ex]
Compound&KLL,&KLL,&Compound&KLL,&KLL,\\[-0.9ex]
 &eV&eV&&eV&eV\\
\hline \hline
Mg(OH)$_{2}$                  &503&-7&BeO                           &510&0\\
SrTiO$_{3}$                   &503&-7&MgO                           &503&-7\\
$\alpha$-Al$_{2}$O$_{3}$      &508&-2&Al(OH)$_{3}$                  &511&1\\
SiO$_{2}$                     &507&-3&Ca(OH)$_{2}$                  &511&1\\
Ca(OH)$_{x}$                  &512&2&CaO                           &509&-1\\
MnO$_{2}$                     &515.1&5.1&FeO                           &510&0\\
Fe$_{2}$O$_{3}$               &508&-2&NiOOH                         &511&1\\
Ni$_{2}$O$_{3}$               &512&2&CuO                           &502&-8\\
Cu$_{2}$O                     &509&-1&Y$_{2}$O$_{3}$                &507&-3\\
InPO$_{x}$                    &508&-2&Sb$_{2}$O$_{5}$               &507&-3\\
Nd$_{2}$O$_{3}$               &511&1&Dy$_{2}$O$_{3}$               &510&0\\
Tb$_{2}$O$_{3}$               &510&0&Tm$_{2}$O$_{3}$               &505&-5\\
Lu$_{2}$O$_{3}$               &510&0&HfO$_{x}$                     &510.8&0.8\\
PbO                           &513&3&SiN$_{x}$O$_{y}$              &508&-2\\
LiNbO$_{3}$                   &512&2&BaTiO$_{3}$                   &509&-1\\
$\gamma$-2CaOSiO$_{2}$        &506&-4&$^\dag$3CaOSiO$_{2}$                 &504&-6\\
$\beta$-2CaOSiO$_{2}$         &510&0&$^\dag$3CaOSiO$_{2}$&511&1\\
CaOSiO$_{5}$0.75TiO           &512&2&$^\dag$CaOSiO$_{5}$0.5TiO&509&-1\\
CaO2Al$_{2}$O$_{3}$           &506&-4&$^\dag$Ca$_{3}$Al$_{2}$O$_{6}$&508&-2\\
$^\dag$3CaOAl$_{2}$O$_{3}$                     &513&3&$^{\dag\dag}$3CaOAl$_{2}$O$_{3}$&516&6\\
12CaO7Al$_{2}$O$_{3}$                   &513&3&2CaOFeO                      &509&-1\\
4CaOAl$_{2}$O$_{3}$Fe$_{2}$O$_{3}$      &513&3&KAl$_{3}$Si$_{3}$O$_{12}$     &502&-8\\
\hline %
\multicolumn{6}{l}{$^\dag$After hydration.}\\
\multicolumn{6}{l}{$^\dag$$^\dag$Prior to hydration.}
\end{tabular}\label{tab3}}
\end{table}

\begin{figure*}
\centering
\includegraphics[width=16 cm]{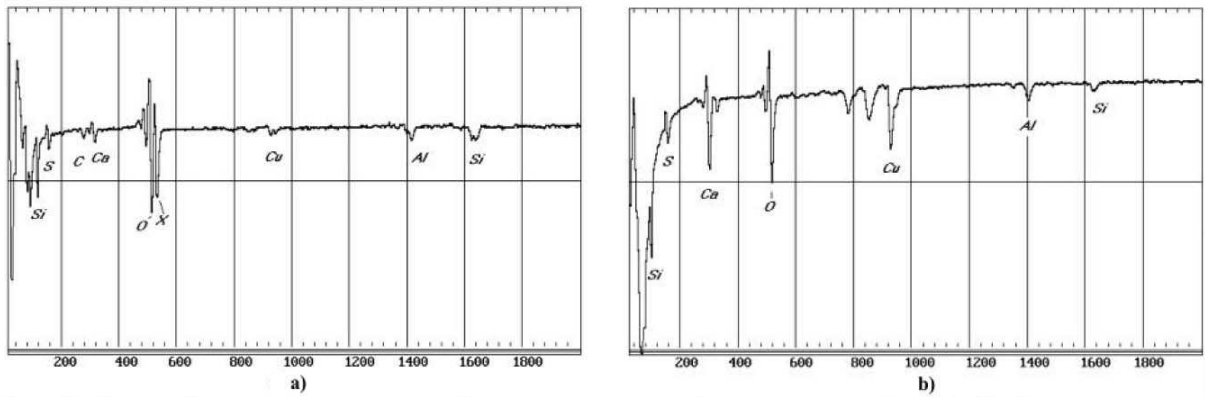}
\caption{Typical Auger-spectrum containing the unidentifiable peak
with an energy of 527\,eV (a) and the spectrum registered on the
same place in two months (b).\label{Auger_5.eps}}
\end{figure*}

As for the stability of this 527-eV peak, we mention a few points.
Its behavior under the action of the electron beam of the probe is
such that, upon the repeated registration of the Auger-spectrum
which was measured at once after the previous one, we failed to
observe it though its intensity was significant upon the first
registration. Unfortunately, the Auger-spectra illustrating this
fact were observed by an operator in the scanning regime and were
not registered as those of no interest at that moment. However, in
two months, we returned to that particle, on which the intensity
of the 527-eV peak was maximum, and studied again the composition
of its surface. Both relevant Auger-spectra are given in
Fig.~\ref{Auger_5.eps}. They also demonstrate the fact that the
content of Ca has considerably increased.

As in the previous case, we tried to find the relevant
low-intensity peaks for the unidentifiable 527-eV peak. However,
we revealed no unidentifiable accompanying peaks upon the
registration of Auger-spectra in a wide energy range and with
large exposure duration on those places of specimens, where we
registered the 527-eV peak.

\begin{figure*}
\centering
\includegraphics[width=16 cm]{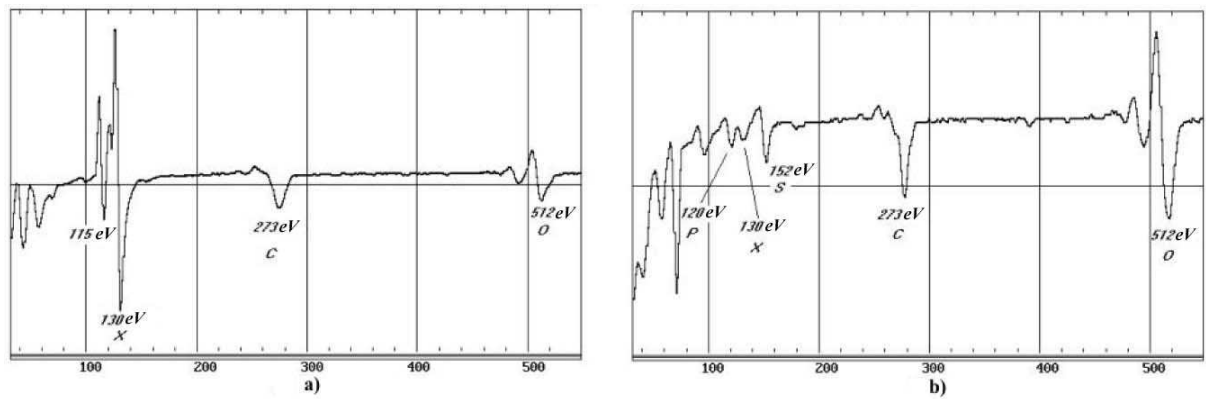}
\caption{Fragment of the Auger-spectrum containing the
unidentifiable doublet of peaks with energies of 115 and 130\,eV
(a) and the fragment of the Auger-spectrum containing these peaks
on the repeated registration (b).\label{Auger_6.eps}}
\end{figure*}

The fourth unidentifiable peak or, more correctly, a doublet of
peaks was registered only on one specimen. The main peak of the
doublet had energy of 130\,eV, and the energy of the less
pronounced peak was 115\,eV. A fragment of the Auger-spectrum
containing these peaks is given in Fig.~\ref{Auger_6.eps}, a.
These peaks were registered only in 2 Auger electron spectra (see
Table~\ref{tab1}). In both cases, the peaks were found on small
($5\dots10\,\mu$m in diameter) light particles. The energy
position of these peaks is quite exactly reproduced, but their
intensity varies from a huge to insignificant one. Their typical
chemical environment consists of Al, O, C, N, S, P, Cl, Cu, Sn,
and Ce.

 As for the stability of these peaks, we say, firstly, a few
words about how they were discovered. At first, after the
Auger-spectrum registration, the peaks were erroneously
interpreted as the Auger-peaks of yttrium from series MNN, because
their energies well coincide. However, we were disturbed by the
following circumstance: the registered peaks had high intensity,
the Auger-spectrum belonged to the energy interval from 30 to
3000\,eV with high exposure duration, but the brightest Auger-peak
of yttrium from series LMM with an energy of 1746\,eV was absent.
In this situation, the natural question arose about whether the
intensity of the 1746-eV Auger-peak of yttrium from series LMM
exceeds the background fluctuations under the conditions of
registration of the Auger-spectrum. To solve this problem, we
registered the standard Auger-spectrum of yttrium being in its
simple compound Y$_{2}$O$_{3}$ under the same conditions. As a
result, we established that the intensities of the Auger-peaks of
yttrium with energies of 127 and 1746\,eV are related
approximately as 4\,:\,1. This means that if, in the situation
under discussion, the low-energy doublet of peaks would belong to
Y, its 1746-eV Auger-peak from series LMM must considerably exceed
the background fluctuations, because the intensity of the
registered doublet was very large (Fig.~\ref{Auger_6.eps}, a).

To verify that the observed doublet of peaks does not correspond
to yttrium, we tried again to register it and to study its
behavior and the spreading area. By returning in several days on
that particle where it was observed, we failed to find the
doublet. While studying the tremendous number of similar
particles, we sometimes succeeded to observe the doublet in the
scanning regime. However, in all the cases, its intensity was
considerably less than that on the first particle, and the doublet
was already absent when we made registration with a large exposure
duration except for one case. The fragment of the Auger-spectrum,
where it was registered for the second time, is presented in
Fig.~\ref{Auger_6.eps}, b. The unidentifiable doublet is also
characterized by the fact that it has no accompanying
low-intensity peaks in the energy range $30\dots3000$\,eV. This
conclusion follows from their absence in the spectrum where the
unidentifiable doublet was registered with high intensity for a
large exposure duration.

\begin{figure*}
\centering
\includegraphics[width=16 cm]{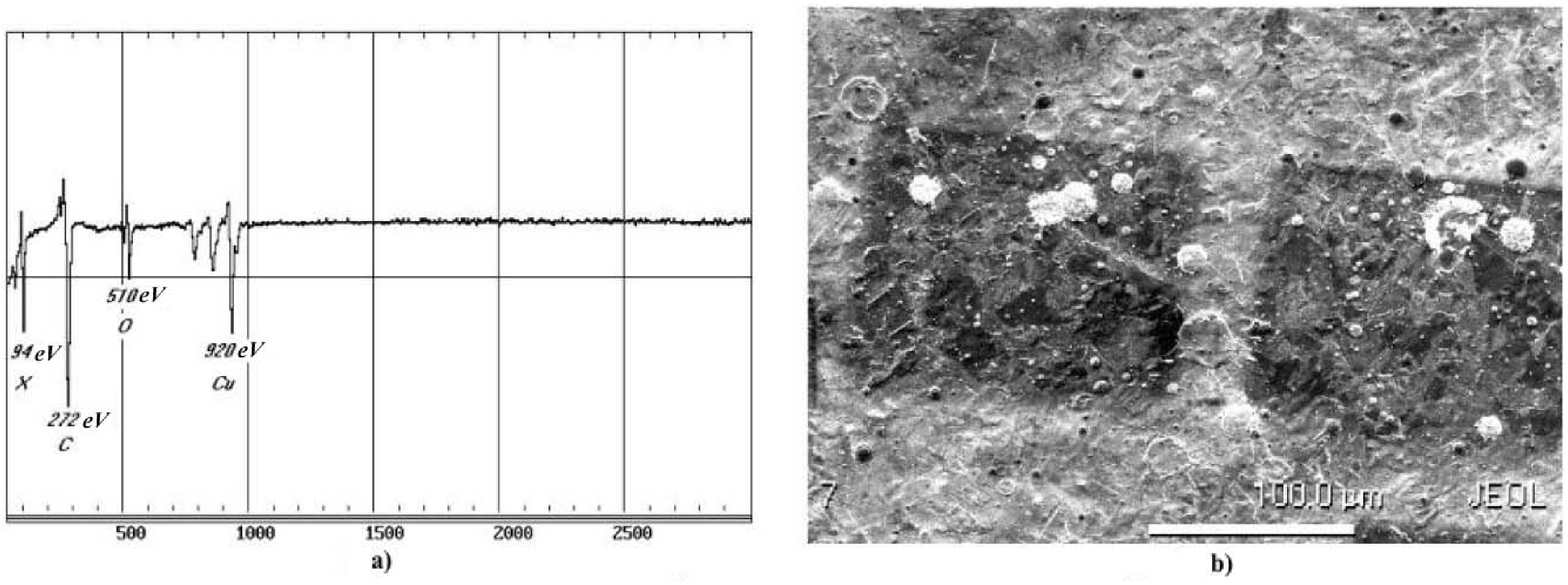}
\caption{Typical Auger-spectrum containing the unidentifiable
94-eV peak (a); the peak was registered on light particles in
craters formed by ion etching (b).\label{Auger_7.eps}}
\end{figure*}

The fifth unidentifiable peak with an energy of 94\,eV was
registered on 2 specimens in 10 Auger-spectra (Table~\ref{tab1}).
One of these spectra containing the peak is shown in
Fig.~\ref{Auger_7.eps}, a. In all the registered spectra, its
intensity was significant and its energy position was quite
exactly reproduced. Its chemical environment was the same in all
cases and rather scanty: C, O, and Cu. It is typical that its
environment did not contain even S, Cl, and N which are usually
present in all spectra of the specimens under study. The peak was
relatively stable in time and under the action of the electron
beam of the probe.

 We now discuss the spreading area of the unidentifiable 94-eV
peak. We studied twice the specimen, where it was registered. At
first, all the specimen surface and all its details were
thoroughly investigated, but we did not find the peak. For the
second time, the same specimen was investigated approximately in a
month after secondary ion mass-spectroscopy study. On the specimen
surface, we observed a significant number of craters of about
$1\,\mu$m in depth and $250\times250\,\mu$m in area which appeared
after the ion etching of the ion gun of a mass-spectrometer of
secondary ions. On the bottom of craters and on the specimen
surface, we observed many circular flattened light particles of
$10\dots40\,\mu$m in diameter (Fig.~\ref{Auger_7.eps}, b). These
particles contain always the significant amount of Pb and
frequently of Si. We note that the Auger-spectra of both elements
in the range $92\dots94$\,eV include rather bright low-energy
Auger-peaks. However, there are also high-energy Auger-peaks with
energies of 2187 and 1619\,eV for Pb and Si, respectively. But on
some parts of the mentioned particles and only on those which were
located on the crater bottom rather than on the specimen surface,
we registered the Auger-spectra containing the 94-eV peak without
Auger-peaks with energies of 2187 and 1619\,eV. In this case, the
intensity of the 94-eV peak was so large that if it would belong
to Pb or Si, the Auger-spectra of both elements must contain also
high-energy Auger-peaks. Since the latter were absent, we refer
the 94-eV peak to unidentifiable ones. This peak is observed quite
frequently, and the number of its registrations can be made as
large as one likes. We note also that the 94-eV peak has no
accompanying low-intensity unidentifiable peaks in the energy
range $30\dots3000$\,eV (see Fig.~\ref{Auger_7.eps}, a).

\begin{figure}[h]
\centering
\includegraphics[width=8 cm]{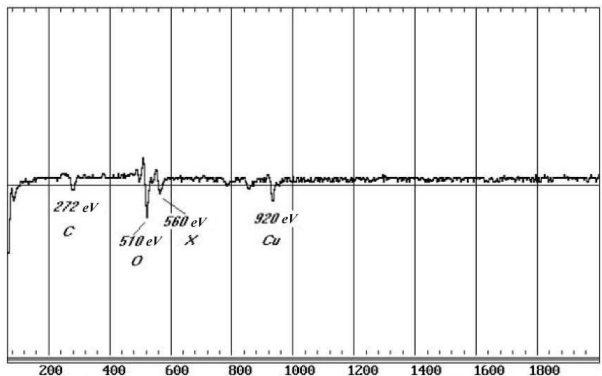}
\caption{Auger-spectrum containing the unidentifiable 560-eV peak
(Cs).\label{Auger_8.eps}}
\end{figure}

The following peak can be referred to unidentifiable ones only
with some reservations. It was registered only on 1 specimen and
only in 1 Auger-spectrum (Table~\ref{tab1}). Its energy position
can be estimated as $559\dots562$\,eV. The Auger-spectrum
containing it is shown in Fig.~\ref{Auger_8.eps}. It was
registered on a small light particle. Like the previous peak, it
had a scanty chemical environment: C, O, and Cu. Its stability is
characterized by the fact that it was not found in the repeated
registration.

As for reservations, they are the following. On the one hand, the
energy position of the peak can be referred to the Auger-peak of
Cs from series MNN with an energy of 563\,eV due to the closeness
of their energies. On the other hand, it is obvious that the
coincidence or the closeness of the energies of both peaks is not
sufficient for such an identification. It is also necessary that
the peaks of other series coincide by energies. However, the other
Auger-peaks of Cs have low intensities as compared to its
analytical Auger-peak (563\,eV), and the spectrum under
consideration was registered under such conditions that the former
can to be not registered against the background (see
Fig.~\ref{Auger_8.eps}). In other words, we have no sufficient
reasons for to say that this peak belongs to Cs, on the one hand.
But, on the other hand, this situation cannot be omitted. In any
case, the peak behavior cannot be referred to the ordinary one.

By summarizing the representation of the derived experimental
results, we note that one of the variants of the interpretation of
the observed unidentifiable Auger-peaks which were not referred to
artifacts is the assumption of their affiliation to long-lived
transuranium elements, which is natural in the context of the
present work. Of interest is the estimation of the atomic numbers
of chemical elements, to which at least some indicated Auger-peaks
can be referred: e.g., the unidentifiable Auger-peaks with
energies of 172, 527, and 1096\,eV.

One of the methods of estimation of the atomic numbers of chemical
elements is the extrapolation based on the Moseley law~\cite{13}
which gives the dependence of the energy of a definite atom level
or the energy of an X-ray line from a specific atom series on the
atomic number. In our situation, the Moseley law can be written%
\begin{equation}%
Z = k E^{1/2}+Z_0,%
\label{Eq1}
\end{equation}%
where $k$ and $Z_{0}$ are the constants in the limits of one
series of Auger-peaks and $Z$ is the atomic number of a chemical
element whose Auger-peak from the series under consideration has
the energy $E$.

The data on the energies of the Auger-peaks of chemical elements
can be found in atlases and catalogs~\cite{10,11,12, 14}. It is
worth to note that we are interested, first of all, in the
energies of the Auger-peaks of chemical elements with large atomic
numbers, but the relevant information available from the
literature is rather scanty. By analyzing the literature data, we
can distinguish only two clear series of bright Auger-peaks of
heavy chemical elements corresponding to the Auger-transitions of
series NOO. Information on them is given in Table~\ref{tab4},
where they are conditionally denoted as Series I and Series II. By
Eq.~\ref{Eq1}, we calculated the values of atomic numbers $Z'$
with $k_\mathrm{I}=2.48$\,eV$^{-1/2}$, $Z_\mathrm{I0}=58.00$, and
$k_\mathrm{II}=4.57$\,eV$^{-1/2}$, $Z_\mathrm{II0}=52.70$ for the
first and second series, respectively. The deviation (see
Table~\ref{tab4}) of the rated value of the atomic number of a
chemical element from its real atomic number, $\triangle Z$
testifies to that Eq.~\ref{Eq1} describes satisfactorily the
series of Auger-peaks under study. We note that a growth of
$\triangle Z$ with decrease in the atomic number is related to the
growing measurement error for the energy position of low-energy
Auger-peaks rather than to the deterioration of the degree of the
used approximation.

\begin{table}[ht]
\centering%
\caption{Approximation by the Moseley law of two series of the
Auger-peaks of heavy chemical elements for Auger-transitions of
the NOO type.\vspace*{1pt}} {\footnotesize
\begin{tabular}{|c|c|c|r|r|c|r|r|}
\hline

$Z$&Element& \multicolumn{3}{|c|}{Series I}& \multicolumn{3}{|c|}{Series II}\\

\cline{3-8}
&&Energy, eV&\multicolumn{1}{|c|}{$Z'$}&\multicolumn{1}{|c|}{$\triangle Z$}&Energy, eV&\multicolumn{1}{|c|}{$Z'$}&\multicolumn{1}{|c|}{$\triangle Z$}\\
\hline \hline
92&U &188&92.00 & 0.00&74&92.01&0.01 \\
90&Th&161&89.47 &-0.53&64&89.26&-0.74\\
83&Bi&101&82.92 &-0.08&44&83.01& 0.01\\
82&Pb&94 &82.04 & 0.04&40&81.60&-0.40\\
81&Tl&84 &80.73 &-0.27&36&80.12&-0.88\\
80&Hg&78 &79.90 &-0.10&  &     &     \\
79&Au&72 &79.04 & 0.04&  &     &     \\
78&Pt&65 &77.99 &-0.01&  &     &     \\
77&Ir&54 &76.22 &-0.78&25&75.55&-1.45\\
76&Os&45 &74.64 &-1.36&  &     &     \\
75&Re&34 &72.46 &-2.54&  &     &     \\
74&W &24 &70.15 &-3.85&  &     &     \\

\hline
\end{tabular}\label{tab4}}
\end{table}

Now, by using the calculated constants ($k_\mathrm{I}$,
$Z_\mathrm{I0}$, $k_\mathrm{II}$, and $Z_\mathrm{II0}$)), we can
estimate the assumed atomic numbers of chemical elements, to which
the nonidentified Auger-peaks with energies of 172, 527, and
1096\,eV would belong in the cases where they refer to Series I
and Series II. The results of these calculations are given in
Table~\ref{tab5}.

\begin{table}[ht]
\centering%
\caption{Extrapolation of the atomic numbers of chemical elements
by two series of Auger-peaks of the NOO transition.\vspace*{1pt}}
{\footnotesize
\begin{tabular}{|c|r|c|r|}

\hline

\multicolumn{2}{|c|}{Series I}& \multicolumn{2}{|c|}{Series II}\\

\hline

Energy, eV& \multicolumn{1}{|c|}{$Z'$}& Energy, eV & \multicolumn{1}{|c|}{$Z'$}\\

\hline \hline

172 & 90.52  &172  &112.63\\
527 & 114.98 &527  &157.71\\
1096& 140.07 &1096 &203.92\\

\hline
\end{tabular}\label{tab5}}
\end{table}

By analyzing the derived data, we should like to make some
remarks. First of all, it is seen from Table~\ref{tab5} that the
Auger-peaks with energies of 172 and 527\,eV could be referred to
two different series of one chemical element with atomic number in
the range $112\dots115$. Moreover, we note as to the 172-eV peak
that its affiliation to a chemical element with atomic number in
the region of $90\dots91$ seems unlikely, because these elements
have brighter Auger-peaks which were absent in the registered
spectra. As to the 1096-eV peak, its affiliation to the considered
series is not so realistic, probably. However, the last remark
does not mean that the mentioned peak cannot be referred in a more
reasonable way to some atomic number of a chemical element. This
peak can be just outside of the considered series, whereas other
series which could be a base for the extrapolation are not clearly
manifested, unfortunately, in the available data~\cite{10,11,12,
14} on the energies of Auger-peaks of heavy elements.

\section{CONCLUSIONS}

By the results of studies of the composition of the laboratory
nucleosynthesis products by local Auger electron spectroscopy, we
arrive at the following conclusions.

\begin{itemize}
\item \emph{While studying the composition of the nucleosynthesis
products on more than one hundred of specimens by AES, we
registered up to several tens of chemical elements in the range
from small to large atomic numbers which did not belong to the
composition of the initial materials of targets and accumulating
screens or, if they were present initially as impurities, exceeded
the total initial content in the mentioned materials by several
orders of magnitude.}

\item \emph{The regular appearance of the new chemical elements as
reaction products, which were absent previously in the composition
of the initial materials of targets and accumulating screens, is
one of the facts testifying to the running of the reactions of
nucleosynthesis in our laboratory setup.}

\item By analyzing the nucleosynthesis products, we registered 6
Auger-peaks (one doublet) with energies of 172, 527, 1096, 94,
560, and 130 (115)\,eV which, on the one hand, are not related to
the known chemical elements and, on the other hand, cannot be
considered as artifacts, in our opinion. These Auger-peaks are
referred by us to the class of unidentifiable ones.

\item As one of the variants of interpretation of the discovered
unidentifiable Auger-peaks, we consider the assumption about their
affiliation to long-lived transuranium elements. By the example of
the unidentifiable Auger-peaks with energies of 172, 527, and
1096\,eV, we performed the extrapolational estimate of atomic
numbers of the chemical elements they could belong to, by using
the Moseley law.
\end{itemize}

\end{document}